\documentstyle[12pt]{article}
\begin{document}
\newcommand{\newc}{\newcommand}

\newc{\be}{\begin{equation}}
\newc{\ee}{\end{equation}}
\newc{\ba}{\begin{eqnarray}}
\newc{\ea}{\end{eqnarray}}
\newc{\bann}{\begin{eqnarray*}}
\newc{\eann}{\end{eqnarray*}}
\newc{\ie}{{\it i.e.}}
\newc{\eg}{{\it eg.}}
\newc{\etc}{{\it etc.}}
\newc{\etal}{{\it et al.}}

\newc{\ra}{\rightarrow}
\newc{\lra}{\leftrightarrow}
\newc{\no}{Nielsen-Olesen }
\newc{\lsim}{\buildrel{<}\over{\sim}}
\newc{\gsim}{\buildrel{>}\over{\sim}}

\begin{titlepage}
\begin{center}
June 1993\hfill
             \hfill
\vskip 1in

{\large \bf
Existence of Double Vortex Solutions
}

\vskip .6in
{\large Leandros Perivolaropoulos}\footnote{E-mail address:
leandros@cfata3.harvard.edu},\footnote{Also, Visiting Scientist,
Department of Physics, Brown University, Providence, R.I. 02912.}\\[.15in]

{\em Division of Theoretical Astrophysics\\
Harvard-Smithsonian Center for Astrophysics\\
60 Garden St.\\
Cambridge, Mass. 02138, USA.}\\[.15in]
\end{center}
\vskip .7in
\begin{abstract}
\noindent
We show analytically and numerically the existence of double vortex solutions
in two-Higgs systems. These solutions are generalizations of the \no vortices
and exist for all values of the parameters in the Lagrangians considered.
 We derive analytically the asymptotic behavior of the solutions and confirm
it numerically by solving the field equations.
 Finally, we show that these solutions can be embedded in realistic theories
like the two-doublet extension of the standard model.
 \end{abstract}

\end{titlepage}

\par
 It has recently been shown\cite{v92,bv92} that the \no vortex
solution\cite{no73,vort} can  be embedded in the standard electroweak model. It
was also
shown\cite{jpv}
 that the resulting electroweak vortices are dynamically (but not
topologically) stable for a finite range of parameters of the
electroweak model Lagrangian.
 This range of parameters however is not included within the limits
allowed by current experiments, assuming the simplest form of the Higgs
sector (one Higgs doublet).

There are recent theoretical and experimental indications\cite{susy} that the
Higgs sector of the standard model may need to be extended to involve
at least two Higgs multiplets for the electroweak symmetry
breaking. This is required, for example, in order to achieve
consistent unification of the gauge couplings on high energy scales\cite{susy}.
The prospect of this possibility raises the question of what (if any)
soliton-like objects exist in these extended, multiple Higgs theories.

Clearly the \no vortex which involves a single Higgs can not be directly
embedded in multiple Higgs systems unless it is generalized first.
Recent studies\cite{ds93} have shown the existence of an extra global symmetry
in realistic two-Higgs doublet models, which breaks during the
electroweak symmetry breaking.
The breaking of this extra symmetry may lead to {\it topologically}
stable embedded {\it double vortices} (the term {\it `double'} here
refers to the number of Higgs fields that wind non-trivially
and not to the topological charge or the flux of the vortices).
It is therefore important to study how can the
\no vortex be generalized to multiple Higgs systems, what are the properties
of the new solutions and how can they be embedded in realistic two doublet
models. These issues consist the focus of the present work.

Previous studies \cite{l92} attempting to generalize the \no solution
have claimed that such generalized solutions exist only for certain values
of the parameters of the two-Higgs Lagrangian. These studies, which did not
make any numerical confirmation of their results, used a very constraining
ansatz for the asymptotic behavior of their candidate solutions which led
to incorrect conclusions about the existence of solutions.

Here, we avoid
the use of {\it any} ansatz and show that vortex solutions exist
for any value of  parameters in the Lagrangian. We derive the asymptotic
behavior of these solutions and point out new features that are not present
in the \no vortices. We also obtain the solutions numerically and confirm
the derived asymptotic behavior.
Finally, we consider the realistic two doublet electroweak model and show that
the obtained double vortex solution can be embedded in this model.

Consider the Abelian-two-Higgs Lagrangian:
\be
{\cal L}= -{1\over 4} F^{\mu \nu} F_{\mu \nu} +
{1\over 2}\vert D_\mu \Phi_1 \vert ^2 +
{1\over 2}\vert D_\mu \Phi_2 \vert ^2
- V_0(\Phi_1, \Phi_2)
\ee
where $\Phi_1$, $\Phi_2$ are complex singlets (the Higgs fields),
$D_\mu = \partial_\mu - ie A_\mu$, $F_{\mu \nu}=\partial_\mu A_\nu-
\partial_\nu A_\nu$ and
\be
V_0(\Phi_1,\Phi_2)={\lambda_1 \over 4} (\vert \Phi_1 \vert ^2 - v_1 ^2)^2 +
{\lambda_2 \over 4} (\vert \Phi_2 \vert ^2 - v_2 ^2)^2 +
{\lambda_3 \over 4} (\vert \Phi_1 \vert ^2 + \vert \Phi_2 \vert ^2 - v_1 ^2 -
v_2 ^2)^2
\ee
Consider now the ansatz:
\ba
\Phi_1 &=&v_1 f_1(r) e^{i\theta}\\
\Phi_2 &=&v_2 f_2(r) e^{i\theta}\\
A_\mu&=&{\hat e}_\theta {v(r)\over {er}}
\ea
Using the ansatz (3)-(5) and a rescaling of the radial coordinate $r\ra
{r\over{\sqrt{v_1 v_2} e}}$ the field equations of motion obtained from (1)
become
\ba
f_1''(r)+{1\over r} f_1'(r)-{{(1-v(r))^2}\over r^2} f_1(r) -
q_1 (f_1(r)^2 - 1) f_1(r)\nonumber \\ - q_3 (f_2(r)^2 - 1) f_1(r) =0\\
f_2''(r)+{1\over r} f_2'(r)-{{(1-v(r))^2}\over r^2} f_2(r) -
q_2 (f_2(r)^2 - 1) f_2(r)\nonumber \\ - q_3 (f_1(r)^2 - 1) f_2(r) =0\\
v''(r)-{1\over r} v'(r) + q_5 (1-v(r))f_1 (r)^2 +  q_6 (1-v(r))f_2 (r)^2 =0
\ea
where
\bann
q_1 = {{\lambda_1 +\lambda_3}\over {e^2}}({v_1 \over v_2}) &\hspace{0.5cm}&
q_2 = {{\lambda_2 +\lambda_3}\over {e^2}}({v_2 \over v_1}) \\
q_3 = {{\lambda_3}\over {e^2}}({v_2 \over v_1}) &\hspace{0.5cm}&
q_4 = {{\lambda_3}\over {e^2}}({v_1 \over v_2}) \\
q_5 = ({v_1 \over v_2}) &\hspace{0.5cm}&
q_4 = ({v_2\over v_1})
\eann
Single valuedness for the fields $\Phi_1$, $\Phi_2$ imply the following
boundary conditions for the system (6)-(8)
\be
f_1 \ra 0 \hspace{0.5cm} f_2 \ra 0 \hspace{0.5cm} v\ra 0 \hspace{0.5cm} {\rm
for} \hspace{0.5cm} r \ra 0
\ee
and
\be
f_1 \ra 1 \hspace{0.5cm} f_2 \ra 1 \hspace{0.5cm} v\ra 1 \hspace{0.5cm} {\rm
for} \hspace{0.5cm} r \ra \infty
\ee
It is easy to see that for $r\ll 1$, $f_1 (r) \sim f_2 (r) \sim r$ and
$v(r) \sim r^2$.

The asymptotic behavior of the fields may also be obtained in a
straightforward way. Define $\delta f_1$, $\delta f_2$ and $\delta v$ by
\be
f_i \ra 1 + \delta f_i \hspace{0.5cm} v \ra 1+ \delta v \hspace{0.5cm} {\rm
for} \hspace{0.5cm} r \gg 1
\ee
where $i=1,2$. By keeping only lowest order terms in $\delta f_1$, $\delta
f_2$,
the system (6)-(8) becomes: \ba
\delta f_1''(r)+{1\over r} \delta f_1'(r)-{{\delta v(r)^2}\over r^2} -
2 q_1 \delta f_1(r) - 2 q_3 \delta f_2(r) &=&0\\
\delta f_2''(r)+{1\over r} \delta f_2'(r)-{{\delta v(r)^2}\over r^2} -
2 q_2 \delta f_2(r) - 2 q_4 \delta f_1(r) &=&0\\
\delta v''(r)-{1\over r} \delta v'(r) + (q_5 +q_6) \delta v(r)  &=&0
\ea
Equation (16) shows that $\delta v$ decays exponentially. This implies that the
system of (12) and (13) becomes mathematically similar to a system of coupled
harmonic oscillators driven by the same `force'. This is a simple
problem but we will go through it in some detail in order to show that there is
always a solution and thus resolve the controversy with Ref. \cite{l92}.
Notice that the term ${{\delta v(r)^2}\over r^2}$ has
been kept in (12) and (13). Neglecting this term, as was done in the original
paper of \no, leads to an asymptotic behavior for the fields, which is
incorrect
for large values of the self coupling when this term dominates. For a proper
treatment of this term in the \no case see Ref. \cite{p93}.

Clearly the solution to (14) may be approximated for $r\gg 1$ by
\be
\delta v = r^\tau e^{-\sigma r} (c_1 ^v +c_2 ^v r^{-1})
\ee
Using this ansatz in (14) leads to
\be
\tau = {1 \over 2} \hspace{0.5cm} {\rm and} \hspace{0.5cm} \sigma = \sqrt{q_5
+q_6}
\ee
This gives the behavior of the `driving force' in the system (12), (13).
The homogeneous solution of this system may now be obtained by using an ansatz
of the form
\be
\delta f_i = c_i ^f r^\alpha e^{-\beta r}
\ee
which leads to
\ba
c_1 ^f (\beta ^2 -2q_1) - c_2^f 2q_3 &=& 0\\
-c_1^f 2q_4 + c_2^f (\beta ^2 -2q_2) &=& 0
\ea
Demanding that this system has a solution implies that the secular determinant
vanishes which in turn leads to
\be
\beta_\pm = [(q_1 +q_2) \pm \sqrt{(q_1-q_2)^2 + 4q_3 q_4}\hspace{3mm}]^{1\over
2}
\ee
The corresponding ratio ${{c_2^f} \over {c_1^f}}$ is
\be
({{c_2^f} \over {c_1^f}})_\pm={{(q_2 -q_1) \pm \sqrt{(q_2-q_1)^2 + 4q_3
q_4}}\over {2q_3}}\equiv \gamma_\pm
\ee
By definition $q_3 q_4 \le q_1 q_2$ (for $\lambda_1,\lambda_2,\lambda_3 \geq
0$) which implies $\beta_\pm \in \Re$ and $\gamma_- < 0 < \gamma_+$.
Therefore, the homogeneous solution to the system (12),(13) is
\ba
\delta f_1^h &=& (c_{1+}^f e^{-\beta_+ r} + c_{1-}^f e^{-\beta_- r})r^\alpha \\
\delta f_2^h &=& (c_{1+}^f \gamma_+ e^{-\beta_+ r} + c_{1-}^f \gamma_-
e^{-\beta_- r})r^\alpha
\ea
where $\alpha$ is a constant that can be determined numerically. The particular
solution of the system is of the form
\be
\delta f_i ^p = d_i^f {{e^{-2 \sigma r}}\over r}
\ee
where $i=1,2$.
 The ansatz (24) leads for $r\gg 1$ to the inhomogeneous system
\ba
d_1 ^f (4\sigma ^2 -2q_1) - d_2^f 2q_3 &=&(c_1^v)^2\\
-d_1^f 2q_4 + d_2^f (4 \sigma ^2 -2q_2) &=& (c_1^v)^2
\ea
where $c_1^v$ is defined in (15).
The constants $d_1^f, d_2^f$ can be obtained by solving the system (25), (26)
which always has a solution.
Therefore, the form of $\delta f_i$, $\delta v$ is:
\ba
\delta f_1 &=& (c_{1+}^f e^{-\beta_+ r} + c_{1-}^f e^{-\beta_- r})r^\alpha +
d_1^f {{e^{-2 \sigma r}}\over r}\\
\delta f_2 &=& (c_{1+}^f \gamma_+ e^{-\beta_+ r} + c_{1-}^f \gamma_-
e^{-\beta_- r})r^\alpha + d_2^f {{e^{-2 \sigma r}}\over r}
\ea
Clearly, the term with the lowest value of the exponent will dominate. We may
therefore distinguish four cases
\begin{description}
\item[1. $\lambda_1 \neq \lambda_2$ (or $v_1 \neq v_2$) and $\beta_- <
2\sigma$.]
 In this case the term proportional to $e^{-\beta_- r}$ dominates and we have
\ba
\delta f_1 &=&  c_{1-}^f e^{-\beta_- r}r^\alpha
 \\
\delta f_2 &=& c_{1-}^f \gamma_- e^{-\beta_- r}r^\alpha \\
\delta v &=& c_1^v e^{-\sigma r} r^{1\over 2}
\ea
Notice that since $\gamma_- <0$, $\delta f_2$ is negative in this case and
therefore $f_2 (r)$ is approaching its asymptotic value (=1) from {\it above}.
This is a new feature which does not appear in the \no solution for {\it any}
value of the parameters. We have checked the validity of this point numerically
(see below).

\item[2. $\lambda_1 \neq \lambda_2$ (or $v_1 \neq v_2$) and $\beta_- >
2\sigma$.] In this case the term proportional to $e^{-2 \sigma r}$ dominates
and
\ba
\delta f_i &=& d_i^f {{e^{-2\sigma r}}\over r}\\
\delta v &=& c_1^v e^{-\sigma r} r^{1\over 2}
\ea
where $i=1,2$.

\item[3. $\lambda_1 = \lambda_2$, $v_1 = v_2$ and $\beta_+ < 2\sigma$.]
This implies by symmetry that $\delta f_1 = \delta f_2$ and therefore, since
$\gamma_- < 0$, we must have $c_{1-}^f =0$.
Therefore
\ba
\delta f_i &=& c_{1+}^f {{e^{-\beta_+ r}}}\\
\delta v &=& c_1^v e^{-\sqrt{2} r} r^{1\over 2}
\ea
since $\gamma_+ = 1$, $\sigma=\sqrt{2}$ for $\lambda_1=\lambda_2$ and
$v_1=v_2$.

\item[4. $\lambda_1 = \lambda_2$, $v_1 = v_2$ and $\beta_+ > 2\sigma$.]
The dominant term now is proportional to $e^{-2\sigma r}$ and $\delta
f_1=\delta f_2$ by symmetry.
Thus
\ba
\delta f_i &=& d_1^f {{e^{-2\sqrt{2} r}}\over r}\\
\delta v &=& c_1^v e^{-\sqrt{2} r} r^{1\over 2}
\ea
\end{description}

We have verified the above cases by plotting
 ${{ln(\vert \delta f_i \vert)}\over r}$ and
${{ln(\vert \delta v \vert)}\over r}$ vs $r$ for several different values of
parameters. In Figures 1 and 2 we show those plots for parameter values
$\lambda_1 = \lambda_2 =\lambda_3 =1$ (case 3) and
$\lambda_1=\lambda_3 =1$, $\lambda_2=2$ (case 1) respectively. The asymptotic
values for all fields are consistent with the corresponding predictions of
exponents. Indeed, for Figure 1 the relevant exponent for $\delta
f_1$, $\delta f_2$ is $\beta_+=2.45$ (case 3) while for Figure 2 the
corresponding exponent is $\beta_-=1.7$ (case 1). In both cases we had
$v_1=v_2$ which implies that the relevant exponent for $\delta v$ is $\sigma =
\sqrt{2}$. Notice also the singularity of ${{ln(\vert \delta f_2 \vert)}\over
r}$ in Figure 2, shown as a sharp minimum due to the discrete nature of the
numerically
constructed plot.
This is due to the fact that $f_2 (r)$ crosses the $f_2 (r)=1$
line (leading to $\delta f_2 =0$) in order to approach its asymptotic value
$f_2
= 1$ from above as predicted by (30) (since $\gamma_- < 0$).

It is instructive to study how does the above analysis get modified in the case
of more general two-Higgs potentials appearing in realistic cases.
Consider for example the potential:
\be
V(\Phi_1,\Phi_2)=V_0(\Phi_1,\Phi_2)+
\lambda_4 (\vert \Phi_1 \vert ^2  \vert \Phi_2 \vert ^2-
\vert \Phi_1^* \Phi_2\vert) + \lambda_5 \vert \Phi_1^* \Phi_2 - v_1 v_2
\vert ^2
\ee
where $V_0(\Phi_1,\Phi_2)$ is defined in (2).
Using the same ansatz as in (3)-(5) we obtain the field equations for $f_1
(r)$,
$f_2(r)$ and $v(r)$ as:
\ba
f_1''(r)+{1\over r} f_1'(r)-{{(1-v(r))^2}\over r^2} f_1(r) -
q_1 (f_1(r)^2 - 1) f_1(r) - \nonumber \\
q_3 (f_2(r)^2 - 1) f_1(r)- p(f_1(r) f_2(r) -1)
f_2(r) =0\\
f_2''(r)+{1\over r} f_2'(r)-{{(1-v(r))^2}\over r^2} f_2(r) -
q_2 (f_2(r)^2 - 1) f_2(r) - \nonumber \\
q_3 (f_1(r)^2 - 1) f_2(r)- p'(f_1(r) f_2(r) -1)
f_1(r)=0\\
v''(r)-{1\over r} v'(r) + q_5 (1-v(r))f_1 (r)^2 +  q_6 (1-v(r))f_2 (r)^2 =0
\ea
where
\be
p = {{\lambda_5}\over {e^2}}({v_2 \over v_1}) \hspace{0.5cm} {\rm and}
\hspace{0.5cm} p' = {{\lambda_5}\over {e^2}}({v_1 \over v_2})
\ee
and $q_i$ as defined previously.
It is easy to show that the equations for $\delta f_i$, $\delta v$ may be
obtained from (12)-(14) by substituting
\ba
2q_1 &\ra& 2q_1 +p \hspace{0.5cm} 2q_2 \ra 2q_2 +p'\\
2q_3 &\ra& 2q_3 +p \hspace{0.5cm} 2q_4 \ra 2q_4 +p'
\ea
and the analysis proceeds in exactly the same way but with different
parameters.

We now show that the obtained solutions can be embedded in realistic
theories. The main steps are similar to those in Ref. \cite{v92} which showed
that the \no vortex is a solution of the equations of motion obtained from the
standard electroweak Lagrangian involving a single Higgs doublet. We therefore
follow the notation of Ref. \cite{v92} with minor modifications.

Consider the bosonic sector of the two-Higgs doublet electroweak Lagrangian:
\be
 L= L_W+ L_B+L_{\Phi_1,\Phi_2}-V(\Phi_1,\Phi_2)
\ee
with
\ba
L_W &=& -{1\over 4} G_{\mu \nu a} G^{\mu \nu a}\\
L_B &=& -{1\over 4} F_{B\mu \nu} F^{B\mu \nu} \\
L_{\Phi_1,\Phi_2}&=& \sum_{i=1}^2 \vert D_\lambda \Phi_i \vert ^2 =
\sum_{i=1}^2 \vert (\partial_\lambda -{1\over 2} i g \tau^a W_\lambda^a
-{1\over 2} i g' B_\lambda) \Phi_i\vert^2
\ea
where $G_{\mu \nu a}$, $F_{B\mu \nu}$  are the usual $SU(2)$ and $U(1)_Y$ gauge
tensors and $W_\lambda ^a$, $B_\lambda$ are the corresponding gauge fields.
Also $V(\Phi_1,\Phi_2)$ is given by (38) with $\Phi_1$, $\Phi_2$ being complex
doublets.

The field equations obtained from (45) are
\ba
\partial_\nu G^{\mu \nu a} - g \varepsilon^{abc} G^{\mu \nu b} W_\nu^c &=&
\sum_{i=1}^2 {i\over 2} g (\Phi^\dagger_i \tau^a D^\mu \Phi_i -
(D^\mu \Phi_i)^\dagger \tau^a \Phi_i) \\
\partial_\nu F^{\mu \nu}&=& \sum_{i=1}^2 {i\over 2} g' (\Phi^\dagger_i D^\mu
\Phi_i - (D^\mu \Phi_i)^\dagger \Phi_i) \\
D_\mu D^\mu \Phi_i &=& -{{\delta V(\Phi_1,\Phi_2)}\over {\delta
\Phi^\dagger_i}}  \ea
which are almost identical to the ones shown in Ref. \cite{v92} apart from the
two-Higgs potential term. Define now $e$, $\theta_w$, $Z^\mu$ and $A^\mu$ as
usual \ie
\ba
g&\equiv& e \cos\theta_w \hspace{0.5cm} g'\equiv e \sin\theta_w\\
Z^\mu &\equiv& \cos\theta_w W^{\mu 3}-\sin\theta_w B^\mu \\
A^\mu &\equiv&\sin\theta_w W^{\mu 3} +\cos\theta_w B^\mu
\ea
and consider the ansatz
\ba
A^\mu&=&W^{\mu 1}=W^{\mu 2}=0 \\
Z^\mu&=&{\hat e}_\theta {{v(r)}\over {e r}}\\
\Phi_1&=&v_1 e^{i\theta} f_1(r) \left( \begin{array}{c} 0 \\ 1 \end{array}
\right)  \\
\Phi_2&=&v_2 e^{i\theta} f_2(r) \left( \begin{array}{c} 0 \\ 1 \end{array}
\right)
\ea
It is straightforward to show that the ansatz (55)-(58) substituted in
(49)-(51) leads to the equations (39)-(41) for $f_1 (r)$, $f_2 (r)$ and $v(r)$.
Therefore, the double vortex solution is also a solution of the field equations
in the two Higgs doublet electroweak model.

It has been argued\cite{ds93} that a certain type of the embedded double vortex
is topologically stable due to an extra global symmetry that amounts to the
freedom of independent $U(1)$ phase transformations of the two doublets. The
breaking of this symmetry during the electroweak transition leads to stability
of the embedded double vortices. In the Lagrangians we have considered however,
the presence of the $\lambda_5$ term does not allow such global transformations
and it explicitly breaks the extra global symmetry. Even in the absence of this
term the investigation of the stability of the embedded vortices needs special
attention. In fact, instabilities may be present even in the topological case
towards repulsion of the centers of the two vortices. This is similar to the
\no n-vortex (n being the winding number) which even though topological is
unstable towards decay to n 1-vortices for large self coupling. In the case of
double vortices it would cost logarithmically infinite energy to place the two
centers infinite distance apart. This is due to the fact that the gauge field
can not effectively screen the angular gradient energy of the two centers when
they are separated. This however leaves open the possibility of repulsion at
small distances which would be favoured by the two-Higgs potential term. These
issues of stability are currently under investigation.

\bigskip

{\bf Acknowledgements}

\noindent
I am grateful to T. Vachaspati for
interesting discussions and for pointing out the problem.
I would also like to thank S. Garcia for his help with the numerics.
 This work was supported by a CfA Postdoctoral Fellowship.
\vskip 1cm
\newpage
\centerline{\large \bf Figure Captions}

{\bf Figure 1:}
The dependence of $ln({\vert x \vert})\over r$ on $r$ for
$x=\delta f_1$, $x=\delta f_2$ and $x=\delta v$ with
$\lambda_1 = \lambda_2=\lambda_3=1$, $v_1 = v_2$ (case 3). The predicted
asymptotic values are $\beta_+=2.45$ for $\delta f_1$, $\delta f_2$ and
$\sigma=\sqrt{2}$ for $\delta v$. Clearly, the numerically obtained asymptotic
values are in good agreement with the predicted ones.

{\bf Figure 2:}
Same as Fig. 1 with
$\lambda_1 = \lambda_3=1$, $\lambda_2=2$, $v_1 = v_2$ (case 1). The predicted
asymptotic values are $\beta_-=1.7$ for $\delta f_1$, $\delta f_2$ and
$\sigma=\sqrt{2}$ for $\delta v$.

\end{document}